\documentclass[aps,pra,twocolumn,showpacs,preprintnumbers,amsmath,amssymb,superscriptaddress,10pt,nofootinbib]{revtex4-1}

\usepackage{amsmath,amssymb}
\usepackage{bm}
\usepackage[utf8]{inputenc} 
\usepackage{dcolumn}
\usepackage{graphicx}
\usepackage{SIunits}
\usepackage{ae}
\usepackage{color} 

\newcommand{\diff}[0]{\text{d}}

\newcommand{\im}{\mathrm{Im}\,}

\newcommand{\vek}[1]{\boldsymbol{\mathbf{#1}}}
\newcommand{\vekh}[1]{\hat{{\boldsymbol{\mathbf{#1}}}}}
\renewcommand{\div}[0]{\nabla\cdot\vek}
\newcommand{\grad}{\nabla}
\newcommand{\curl}{\nabla\times\vek}

\newcommand{\be}{\begin{equation}}
\newcommand{\ee}{\end{equation}}
\newcommand{\ba}{\begin{align}}
\newcommand{\ea}{\end{align}}
\newcommand{\e}[1]{\text{e}^{#1}}

\newcommand{\TE}{\text{\tiny TE}}
\newcommand{\TM}{\text{\tiny TM}}

\begin{document}

\title{Fresnel's equations in statics and quasistatics}

\author{Johannes Skaar}
\affiliation{Department of Technology Systems, University of Oslo, Box 70, NO-2027 Kjeller, Norway}
\email{johannes.skaar@its.uio.no}


\begin{abstract}
Fresnel's equations describe reflection and transmission of electromagnetic waves at an interface between two media. It turns out that these equations can be used in quasistatics and even statics, for example to straightforwardly calculate magnetic forces between a permanent magnet and a bulk medium. This leads to a generalization of the classical image method.
\end{abstract}

\maketitle

\section{Introduction}
Can electrostatic or magnetostatic phenomena be described by Fresnel's equations for reflection and transmission at an interface? For example, if a permanent magnet is located in the vicinity of a semi-infinite medium with permeability $\mu$, can we calculate the attraction using Fresnel's equations? The answer is not obvious, as the eigenmodes for frequency $\omega>0$ are transverse electromagnetic waves, while for $\omega=0$ the eigenmodes are longitudinal. Moreover, a crucial step in the textbook derivation \cite{landau_lifshitz_edcm, saleh} of the Fresnel equations is to relate the electric and magnetic fields. Derivations limited to propagating, plane waves use the constant ratio between the electric and magnetic field amplitudes, equal to the wave impedance. Clearly these derivations do not apply in statics, where the electric and magnetic fields are decoupled. Alternatively, one may use possibly complex wavevectors, and express the magnetic field $\vek H$ from the electric field $\vek E$ using Faraday's law,
\be\label{faraday}
\omega\mu\vek H=\vek k\times\vek E,
\ee
for a plane wave with wavevector $\vek k$. Here, we have the complications that in statics, $\omega=0$, and also, the electric field is longitudinal, $\vek k\times\vek E=0$. 

Despite these challenges, we will prove that the Fresnel equations apply even in statics (Sec. \ref{sec:statics}). For the special case with a point charge in the vicinity of a conductor plane or dielectric half-space, the Fresnel equations lead to the classical image method from electrostatics. Similarly, the Fresnel equations give the image method for calculating the fields when a magnetic source is located in the vicinity of a magnetic medium.

In quasistatics the Fresnel equations turn out to be useful to calculate the interaction between a time-varying magnetic source and a conducting medium (Sec. \ref{sec:quasistatics}). Also here, the Fresnel equations lead to an image method, but with the reflection coefficient as a spatial low-pass filter, acting on the field from the image. This gives valuable information about the strength of the interaction as a function of spatial frequency or characteristic size of the source.

We also consider the cases where a static electric or magnetic field source is located in the vicinity of a moving medium. Also here we obtain relatively simple image methods for calculating the electromagnetic field, even for relativistic velocities. Dependent on the orientation of the source, certain interesting effects arise. For example, when describing electric reflection from magnetic sources (or vice versa), the image is Hilbert transformed. With the filtered image model, we can also confirm the recently reported nonreciprocity associated with moving media \cite{prat-camps2018}.

The structure of the paper is as follows. In Sec. \ref{sec:dynamics} the Fresnel equations of electrodynamics, and their conventional proof, are reviewed. In the last part of the section, another proof is given, which will turn out to be valid in statics as well. In Sec. \ref{sec:statics} we consider a Fourier decomposition of the static fields, and prove that the Fresnel equations lead to the image methods from electrostatics and magnetostatics. In Sec. \ref{sec:quasistatics} we use the Fresnel equations in quasistatics, to describe interaction between e.g. a time-varying magnetic source and a conducting medium. Finally, we consider a moving medium in Sec. \ref{sec:moving}.

\section{Fresnel's equations in electrodynamics}\label{sec:dynamics}
Before going to statics, we review the Fresnel equations in electrodynamics \cite{landau_lifshitz_edcm}.
\begin{figure}[]
\begin{center}
\includegraphics[width=7.5cm]{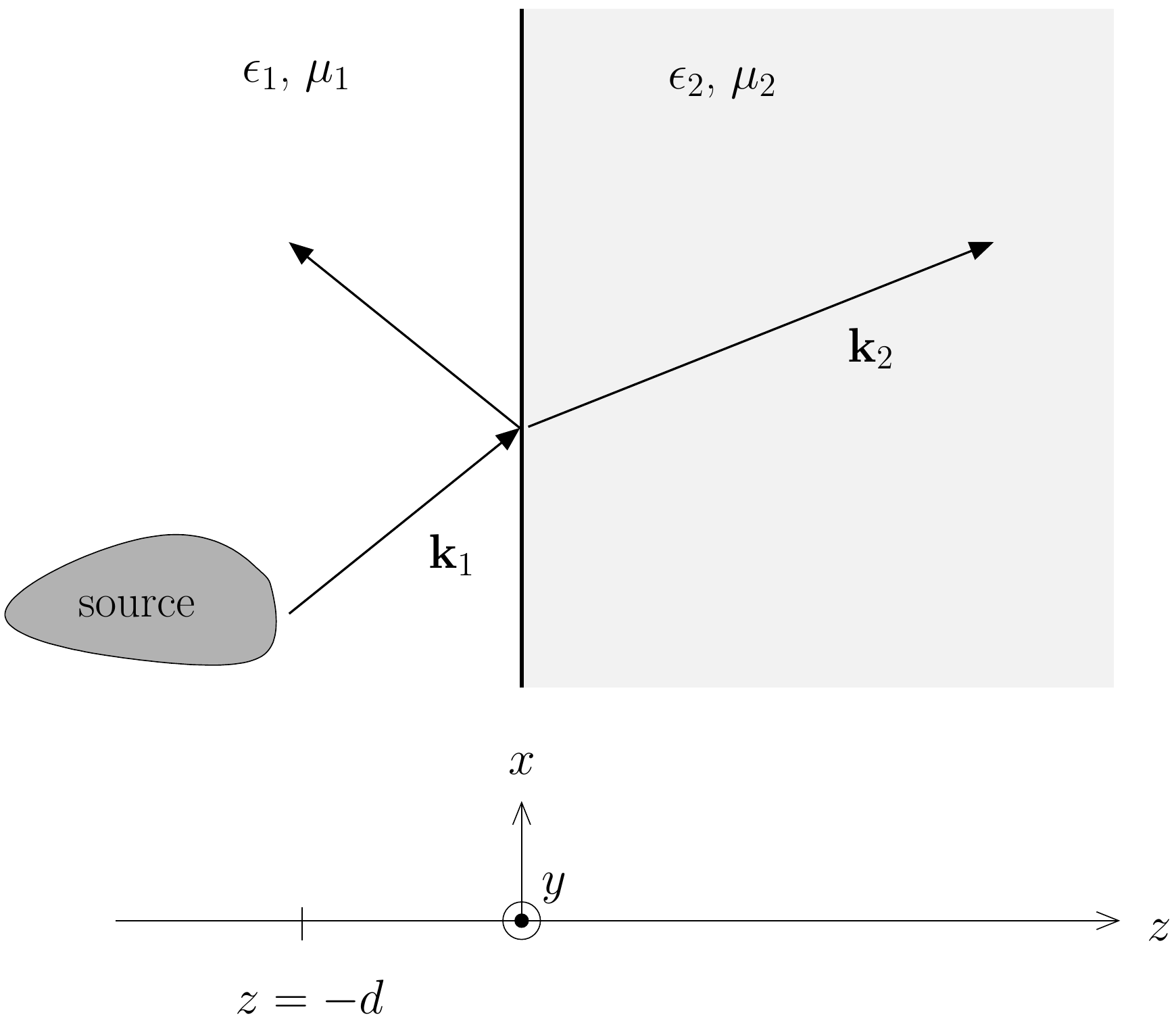} 
\caption{The Fresnel setup. A source is located at least $d$ to the left of the boundary. The source produces a field which can be decomposed into plane waves. One of them has incident wavevector $\vek k_1$.}\label{fig:setup} 
\end{center}
\end{figure}
We consider a setup as depicted in Fig. \ref{fig:setup}. For $z<0$ we have a medium with permittivity $\epsilon_1$ and permeability $\mu_1$, while for $z>0$ we have $\epsilon_2$ and $\mu_2$. The media can be dispersive, but are assumed to be linear, isotropic, homogeneous, time-shift invariant, and passive\footnote{For gain media, there is a complication to identify the correct sign of the longitudinal wavevector based on causality \cite{skaar06,nistad08,kolokolov99,haagenvik15}.}. Moreover, the media are assumed to be spatially nondispersive, in the sense that the permittivity and permeability are local and describe all induced charges and currents. A source produces an electromagnetic field which can be expanded into monochromatic, plane waves. These waves will have their electric field perpendicular (TE) or parallel (TM) to the plane of incidence. Considering an incident TE wave, the electric field can be written
\be\label{ETEbm}
\vek E = 
\begin{cases}
\vekh y\left(E_{\text i}\e{ik_xx+ik_{1z}z} + E_{\text r}\e{ik_xx-ik_{1z}z}\right), & z<0, \\
\vekh y\, E_{\text t} \e{ik_xx+ik_{2z}z}, & z\geq 0.
\end{cases}
\ee
Here $E_\text i$, $E_\text r$, and $E_\text t$ are the incident, reflected, and transmitted electric field amplitudes, respectively. We have oriented the coordinate system such that the electric field points in the $\vekh y$-direction. The wavevectors can then be expressed as $\vek k_1=(k_x,0,k_{1z})$ (incident), $(k_x,0,-k_{1z})$ (reflected), and $\vek k_2=(k_x,0,k_{2z})$ (transmitted), with the dispersion relations
\begin{subequations}\label{disprel}
	\begin{align}
	k_x^2+k_{1z}^2 &= \epsilon_1\mu_1\omega^2, \\
	k_x^2+k_{2z}^2 &= \epsilon_2\mu_2\omega^2.
	\end{align}
\end{subequations}

The Fresnel equations can be expressed in several different forms. We will use a form which remains valid for lossy media and/or evanescent modes. The standard way to derive the equations is as follows. From \eqref{ETEbm} and continuity of the tangential electric field, we obtain $E_\text i+E_\text r = E_\text t$. Assuming $\omega\neq 0$ and applying \eqref{faraday}, we can find the associated magnetic fields for the incident, reflected and transmitted waves. Continuity of the tangential magnetic field gives $\left(E_\text i - E_\text r\right)k_{1z}/\mu_1 = E_\text t k_{2z}/\mu_2$. Combining the two continuity equations we find 
\begin{subequations}\label{TE}
\begin{align}
r^\TE_E &\equiv \frac{E_\text r}{E_\text i} = \frac{\mu_2k_{1z}-\mu_1k_{2z}}{\mu_2k_{1z}+\mu_1k_{2z}}, \\
t^\TE_E &\equiv \frac{E_\text t}{E_\text i} = \frac{2\mu_2k_{1z}}{\mu_2k_{1z}+\mu_1k_{2z}}.
\end{align}
\end{subequations}
These two coefficients express the reflected and transmitted \emph{electric} field amplitudes relative to the incident \emph{electric} field. 

To express the corresponding ratio between \emph{magnetic} field amplitudes (but still TE polarization), we consider the $\vekh x$- and $\vekh z$-components separately. Using Faraday's law \eqref{faraday}, the incident magnetic field $(H_{\text ix},0,H_{\text iz})$ can be expressed
\begin{subequations}
	\begin{align}
	H_{\text ix} &= -\frac{k_{1z}}{\omega\mu_1} E_\text{i}, \\
	H_{\text iz} &= \frac{k_{x}}{\omega\mu_1} E_\text{i}.
	\end{align}
\end{subequations}
Similarly, we can express the reflected and transmitted magnetic fields from the electric fields. For example, this gives us a transmission coefficient for the $\vekh x$-component of the magnetic field 
\be
t^\TE_{H_x}=\frac{-k_{2z}/\omega\mu_2}{-k_{1z}/\omega\mu_1}\cdot\frac{E_\text t}{E_\text i}= \frac{k_{2z}\mu_1}{k_{1z}\mu_2}t^\TE_E.
\ee
We end up with two sets of Fresnel equations:
\begin{subequations}\label{TEHxz}
	\begin{align}
	r_{H_x}^\TE &= -\frac{\mu_2k_{1z}-\mu_1k_{2z}}{\mu_2k_{1z}+\mu_1k_{2z}}, \\
	r_{H_z}^\TE &= \frac{\mu_2k_{1z}-\mu_1k_{2z}}{\mu_2k_{1z}+\mu_1k_{2z}}, \\
	t_{H_x}^\TE &= \frac{2\mu_1k_{2z}}{\mu_2k_{1z}+\mu_1k_{2z}}, \\
	t_{H_z}^\TE &= \frac{2\mu_1k_{1z}}{\mu_2k_{1z}+\mu_1k_{2z}}.
	\end{align}
\end{subequations}

As evident from the symmetry of the two Maxwell curl equations, the analogous expressions for TM (magnetic field perpendicular to the plane of incidence) can be found by interchanging $\epsilon$ and $\mu$:
\begin{subequations}\label{TMH}
\begin{align}
r^\TM_H &= \frac{\epsilon_2k_{1z}-\epsilon_1k_{2z}}{\epsilon_2k_{1z}+\epsilon_1k_{2z}}, \\
t^\TM_H &= \frac{2\epsilon_2k_{1z}}{\epsilon_2k_{1z}+\epsilon_1k_{2z}},
\end{align}
\end{subequations}
and
\begin{subequations}\label{TMExz}
\begin{align}
r^\TM_{E_x} &= -\frac{\epsilon_2k_{1z}-\epsilon_1k_{2z}}{\epsilon_2k_{1z}+\epsilon_1k_{2z}}, \\
r^\TM_{E_z} &= \frac{\epsilon_2k_{1z}-\epsilon_1k_{2z}}{\epsilon_2k_{1z}+\epsilon_1k_{2z}}, \\
t^\TM_{E_x} &= \frac{2\epsilon_1k_{2z}}{\epsilon_2k_{1z}+\epsilon_1k_{2z}}, \\
t^\TM_{E_z} &= \frac{2\epsilon_1k_{1z}}{\epsilon_2k_{1z}+\epsilon_1k_{2z}}.
\end{align}
\end{subequations}

When deriving the Fresnel equations \eqref{TE}, we used the transverse electromagnetic modes of the system, meaning that $\omega$ was assumed to be nonzero. We also made the explicit assumption $\omega\neq 0$ when applying \eqref{faraday}. An alternative way of proving the Fresnel equations, which turns out to survive the static limit, is the following: From Gauss' law, we have 
\be\label{gauss}
\vek k\cdot\vek E=0
\ee
in the homogeneous media to the left or right of the boundary, away from the source. Eq. \eqref{gauss} is valid for the incident, reflected, and transmitted wave, separately. Note that the Fourier decomposition of the source field at the plane $z=-d$ leads to real $k_x$ and $k_y$ (while $k_z$ as resulting from the dispersion relation may be complex). Orienting coordinates such that $\vek k$ is perpendicular to $\vekh y$, 
\be\label{zxkE}
k_xE_x + k_zE_z = 0.
\ee 

Consider the TM case, where $\vek E$ is in the plane of incidence,
\be\label{TMcond}
\vek E=E_x\vekh x + E_z\vekh z.
\ee
Requiring the tangential electric field to be continuous, we get
\be\label{Econt}
E_{\text{i}x} + r^\TM_{E_x}E_{\text{i}x} = t^\TM_{E_x}E_{\text{i}x},
\ee
where the reflection and transmission coefficients so far are unknown. Rather than expressing continuity of the magnetic field, we require the normal displacement field to be continuous:
\be\label{Dcont}
\epsilon_1 E_{\text{i}z} + \epsilon_1 r^{\TM}_{E_z}E_{\text{i}z} = \epsilon_2 t^{\TM}_{E_z}E_{\text{i}z}.
\ee
From \eqref{zxkE}, the connection between the Fresnel coefficients for the $z$ and $x$ components are
\begin{subequations}\label{zxconn}
\begin{align}
r^\TM_{E_z} &= \frac{E_{\text{r}z}}{E_{\text{i}z}} = \frac{\frac{k_x}{k_{1z}}E_{\text{r}x}}{-\frac{k_x}{k_{1z}}E_{\text{i}x}}=-r^\TM_{E_x}, \\
t^{\TM}_{E_z} &= \frac{E_{\text{t}z}}{E_{\text{i}z}} = \frac{-\frac{k_x}{k_{2z}}E_{\text{t}x}}{-\frac{k_x}{k_{1z}}E_{\text{i}x}}=\frac{k_{1z}}{k_{2z}}t^\TM_{E_x}.
\end{align}
\end{subequations}
Here subscripts ``r'' and ``t'' stand for reflected and transmitted, respectively. Canceling out the incident fields from \eqref{Econt} and \eqref{Dcont}, and using \eqref{zxconn}, we obtain \eqref{TMExz}.

\section{Fresnel's equations in statics}\label{sec:statics}
We now assume static conditions, i.e., $\omega=0$. First we consider an electrostatic source, which is a static charge distribution located at least $d$ to the left of the boundary, see Fig. \ref{fig:setup}. A central point to enable the use of Fresnel's equations, is a plane wave expansion of the field. As detailed in the Appendix, a plane wave expansion is possible in statics as well; however the field is longitudinal. The electrostatic field for $-d<z<0$ can be written as a superposition of fields of the type (see the Appendix):
\begin{align}
\vek E(k_x,k_y) &= (k_x,k_y,k_z) U\e{ik_xx+ik_yy+ik_zz} \nonumber\\
&+ (k_x,k_y,-k_z) V \e{ik_xx+ik_yy-ik_zz}, \label{increflw}
\end{align}
where $k_x$ and $k_y$ are real, and
\be\label{disprels}
k_z=i\sqrt{k_x^2+k_y^2}.
\ee
For $z>0$, the field can be written as a superposition of
\be\label{transmw}
\vek E(k_x,k_y) = (k_x,k_y,k_z) W\e{ik_xx+ik_yy+ik_zz}.
\ee
Here the amplitudes $U$, $V$, and $W$ are arbitrary functions of $k_x$ and $k_y$. Morover, $k_x$ and $k_y$ are the Fourier variables corresponding to $x$ and $y$, respectively (i.e., they are the transversal ``spatial frequencies'' of the source). The quantity $k_z$ describes the $z$-dependence of each Fourier component, and is given by \eqref{disprels} as resulting from Maxwell's equations. Eq. \eqref{disprels} can be seen as the dispersion relation in the static limit. 

To apply the proof of Fresnel's equations from the previous section, we identify the first and second term in \eqref{increflw} as the ``incident wave'' $\vek E_\text{i}$ and ``reflected wave'' $\vek E_\text{r}$, respectively, and \eqref{transmw} as the ``transmitted wave'' $\vek E_\text{t}$. We note that $\vek k\cdot\vek E_\text{i}=0$, $(k_x,k_y,-k_z)\cdot\vek E_\text{r}=0$, and $\vek k\cdot\vek E_\text{t}=0$, since \eqref{disprels} implies $\vek k^2=0$. Furthermore, the ``TM'' condition \eqref{TMcond} is satisfied, if coordinates are rotated such that $\vek k=k_x\vekh x+k_z\vekh z$. Apparently, in statics the label ``TM'' refers to electric sources, while, as we shall see later, ``TE'' refers to magnetic sources. 

We conclude that \eqref{TMExz} applies to the case with an electrostatic source. Since \eqref{disprels} is valid in both media, the equations can be simplified to
\begin{subequations}\label{TMstatics}
\begin{align}
r_E &\equiv \frac{V}{U} = \frac{E_{\text{r}x}}{E_{\text{i}x}} = \frac{E_{\text{r}y}}{E_{\text{i}y}} = - \frac{E_{\text{r}z}}{E_{\text{i}z}} = \frac{\epsilon_1-\epsilon_2}{\epsilon_1+\epsilon_2}, \\
t_E &\equiv \frac{W}{U} = \frac{E_{\text{t}x}}{E_{\text{i}x}} = \frac{E_{\text{t}y}}{E_{\text{i}y}} = \frac{E_{\text{t}z}}{E_{\text{i}z}} = \frac{2\epsilon_1}{\epsilon_1+\epsilon_2}.
\end{align}
\end{subequations}
Note that in statics, there is only a single polarization, as the field is longitudinal. Thus it makes sense to write \eqref{TMstatics} for both $x$ and $y$ directions, not requiring the coordinate system to be rotated such that $k_y=0$.

\begin{figure}[]
\begin{center}
\includegraphics[width=7.5cm]{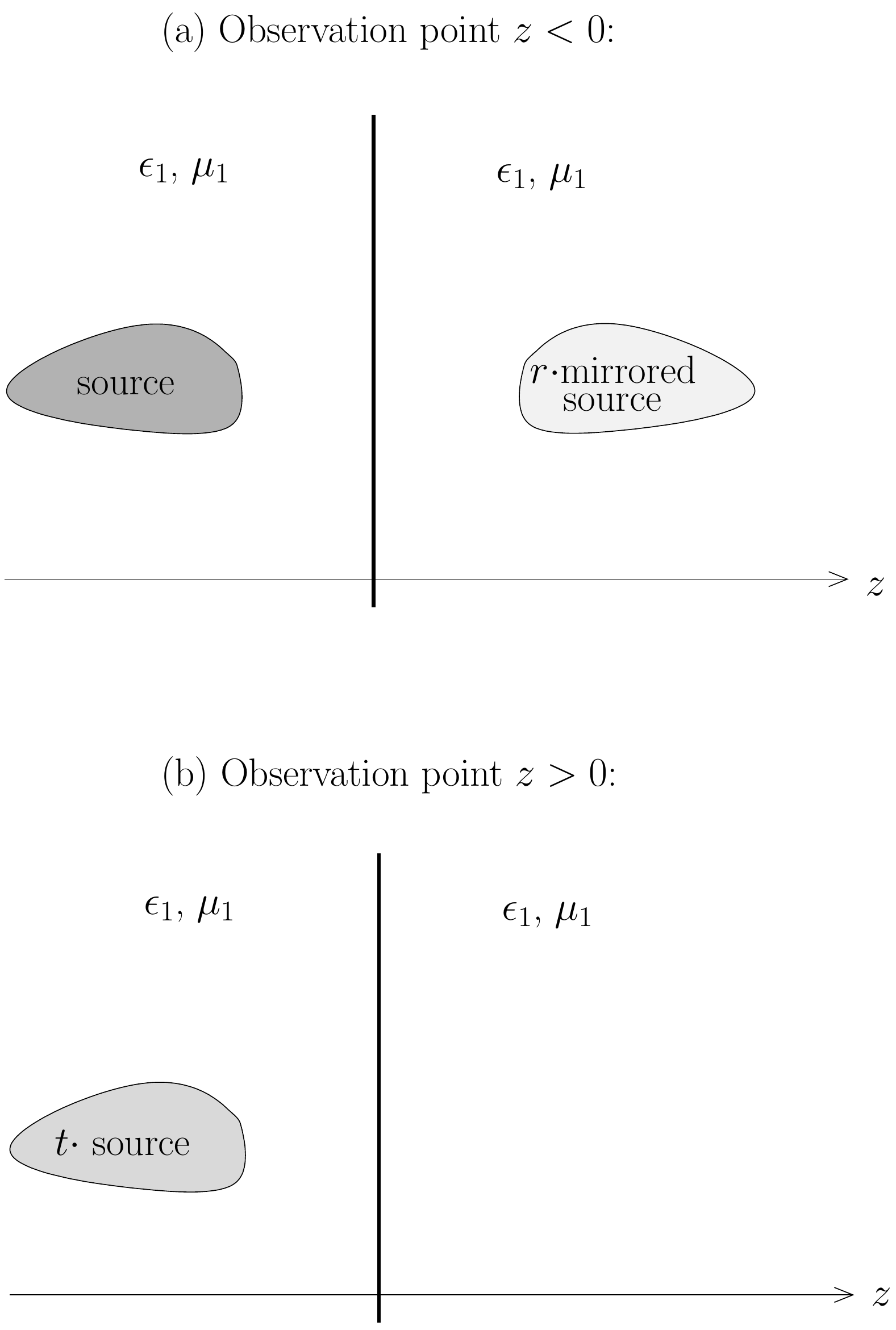} 
\caption{(a) The field for $z<0$ would be obtained if medium 2 is replaced by medium 1, the original source is retained, and there is an image source which is the original source mirrored about the plane $z=0$ and multiplied by $r$. (b) The field for $z>0$ would be obtained if medium 2 is replaced by medium 1, and the original source is multiplied by a factor $t$. In the electrostatic case $r=r_E$ and $t=t_E$. In the magnetostatic case $r=\pm r_H$ and $t=t_H$. The upper sign is used if the source is characterized by a magnetization density, while the lower sign is used when the source is described by a current density.} \label{fig:imageresult} 
\end{center}
\end{figure}

Remarkably, the reflection and transmission coefficients are independent\footnote{For $k_x=k_y=0$ the proof does not apply. However, this 1D special case is trivial in electrostatics; the electric displacement field (in the $\vekh z$-direction) ends up uniform.} of $k_x$ and $k_y$. Therefore, when considering an actual finite-size source, the coefficients can be moved outside the two-dimensional inverse Fourier transform with respect to $k_x$ and $k_y$. Thus the field for $z>0$ would be obtained if we replaced medium 2 by medium 1 (such that medium 1 is everywhere), and replaced the original charge distribution by the exact same charge distribution multiplied by the factor $t_E$ (Fig. \ref{fig:imageresult}b). The field for $z<0$ would be obtained by replacing medium 2 by medium 1, keeping the original charge distribution and inserting an image charge distribution. This new charge distribution must be given by the original charge distribution, mirrored about the plane $z=0$, and multiplied by the factor $r_E$ (Fig. \ref{fig:imageresult}a).

Note that the image charge interpretation above is a generalization of the classical result for a point charge above a dielectric half-space, see e.g. Jackson \cite{jackson}, Chapt. 4.4. In addition it coincides with the image charge method for a point charge above a conducting plane; then we find $r_E=-1$ meaning that the image charge must be exact opposite to the original charge when looking at the field in the area $z<0$, while we have $t_E=0$ which means that the field vanishes for $z>0$.

An identical analysis can be used if the source is a current distribution or permanent magnet producing a magnetic field. Then, away from the source we have $\curl H=0$. Moreover, $\div B=0$ everywhere. For the electrostatic situation treated above, we had $\curl E=0$ and $\div D=0$ away from the source. Thus the analogy $\vek E\to\vek H$ and $\vek D\to\vek B$ makes it possible to express the magnetic field as a superposition of fields of the form
\begin{align}
\vek H(k_x,k_y) &= (k_x,k_y,k_z) S\e{ik_xx+ik_yy+ik_zz} \nonumber\\
&+ (k_x,k_y,-k_z) T \e{ik_xx+ik_yy-ik_zz}, \label{increflwH}
\end{align}
for $z<0$, and
\be
\vek H(k_x,k_y) = (k_x,k_y,k_z) G\e{ik_xx+ik_yy+ik_zz}
\ee
for $z>0$. Furthermore, we can use the Fresnel equations in the same way as above, now using the ``TE'' coefficients
\begin{subequations}\label{TEstatics}
\begin{align}
r_H &\equiv \frac{T}{S} = \frac{H_{\text{r}x}}{H_{\text{i}x}} = \frac{H_{\text{r}y}}{H_{\text{i}y}} = -\frac{H_{\text{r}z}}{H_{\text{i}z}} = \frac{\mu_1-\mu_2}{\mu_1+\mu_2}, \label{TEstaticsr}\\
t_H &\equiv \frac{G}{S} = \frac{H_{\text{t}x}}{H_{\text{i}x}} = \frac{H_{\text{t}y}}{H_{\text{i}y}} = \frac{H_{\text{t}z}}{H_{\text{i}z}} = \frac{2\mu_1}{\mu_1+\mu_2}.
\end{align}
\end{subequations}

Again the reflection and transmission coefficients are independent of $k_x$ and $k_y$. Eqs. \eqref{TEstatics} therefore lead to the following image source interpretation: The magnetic field for $z<0$ is given by the magnetic field from the original source in addition to that from an imaged source, after having replaced medium 2 by medium 1. The image source must provide an imaged field distribution, reflected about the plane $z=0$, and multiplied by $r_H$, according to \eqref{TEstaticsr}. Thus, if the source is a permanent magnet with magnetization $\vek M$, the magnetization of the image source is found by reflecting the source's magnetization about $z=0$ and multiplying by $r_H$. However, if the source is a current distribution, the current must be multiplied by $-r_H$ after reflection. This sign shift can be viewed as a consequence of the current being a polar vector, while $\vek H$ and $\vek M$ are axial. The image source interpretation is a generalization of the well known result from magnetostatics \cite{jackson}.

As an example, consider a bar magnet with magnetization in the $\vekh z$-direction, located a distance $d$ from a magnetic medium with $\mu_2=\infty$. Assuming $\mu_1=\mu_0$ (permeability in vacuum), we have $r_H=-1$. An imaged bar magnet would have magnetization in the $-\vekh z$-direction, but after multiplication with $r_H$ the magnetization points in the $\vekh z$-direction. We therefore get a magnetic attraction equal to the attraction between two identical bar magnets at $2d$ distance.

\section{Fresnel's equations in quasistatics}\label{sec:quasistatics}
In quasistatics, provided 
\begin{subequations}\label{condqs}
\begin{align}
k_x &\gg \omega\sqrt{|\epsilon_1\mu_1|},  \\
k_x &\gg \omega\sqrt{|\epsilon_2\mu_2|},
\end{align} 
\end{subequations}
we have $k_{1z}\approx k_{2z} \approx ik_x$. Therefore, the electrostatic result (with a charge distribution as source) and magnetostatic result (with a current distribution/permanent magnet as source) will also be relevant to the quasistatic situation. An electrostatic source influences a dielectric/conductor medium, while a magnetostatic source influences a magnetic medium. However, in quasistatics the two other combinations are relevant as well; a time-dependent current distribution/magnet as a source and dielectric/conductor media, or a time-dependent charge distribution and  magnetic media. We consider the former combination here.

\begin{figure}[]
\begin{center}
\includegraphics[width=7.7cm]{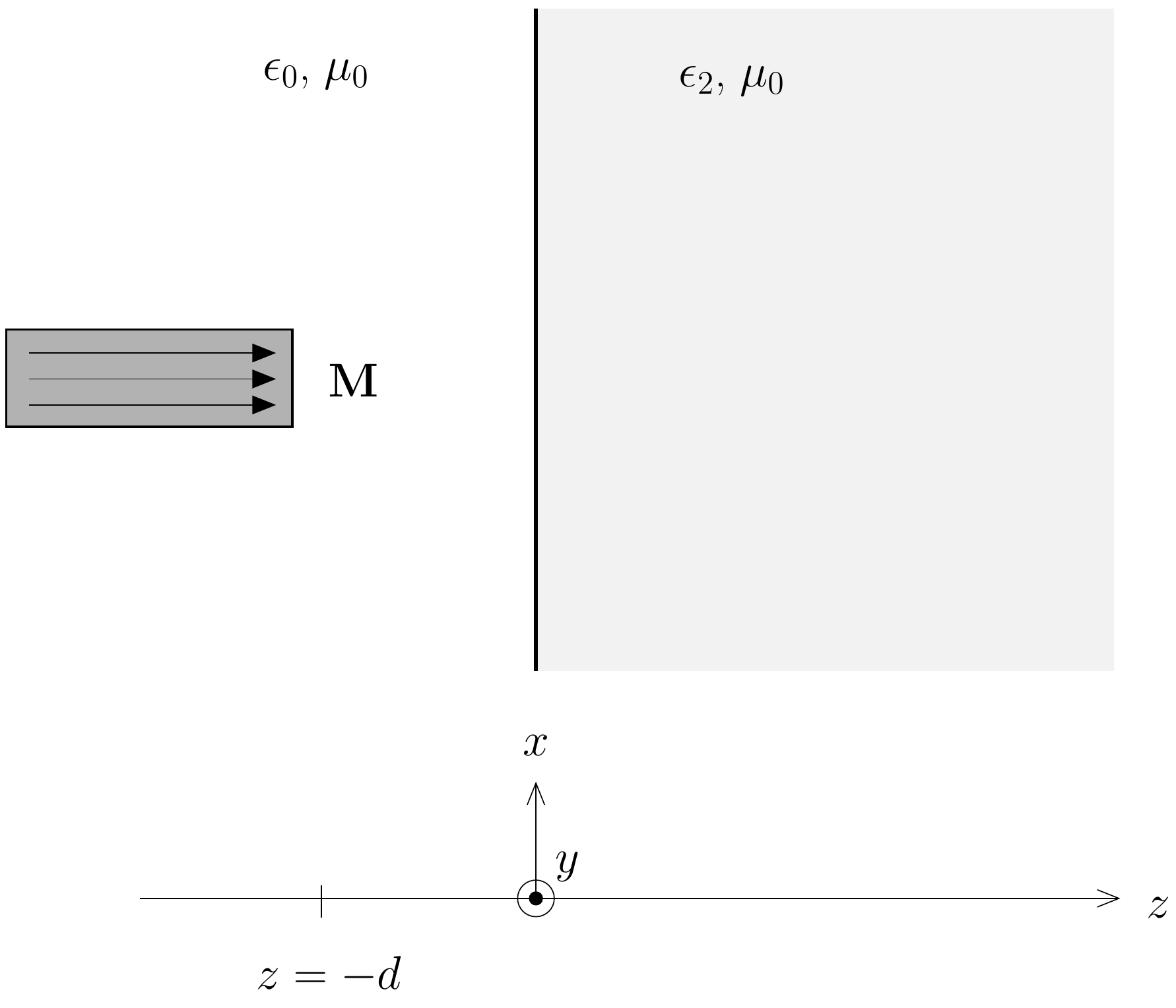} 
\caption{The source is a 2D ``permanent'' magnet with time-dependent magnetization in the $\vekh z$-direction. In practice, the time dependency can be realized by moving the magnet in the $\vekh x$ or $\vekh z$-direction, or by inducing magnetization in a ferromagnetic material using currents in the $\vekh y$-direction. The magnet is infinite and uniform along the $y$-axis. Medium 1 is vacuum, with $\epsilon_1$ and $\mu_1$ equal to the free space values $\epsilon_0$ and $\mu_0$, respectively. Medium 2 is a nonmagnetic conductor with complex permittivity $\epsilon_2$.}\label{fig:barmagnet} 
\end{center}
\end{figure}

\begin{figure}[]
\begin{center}
\includegraphics[width=7.2cm]{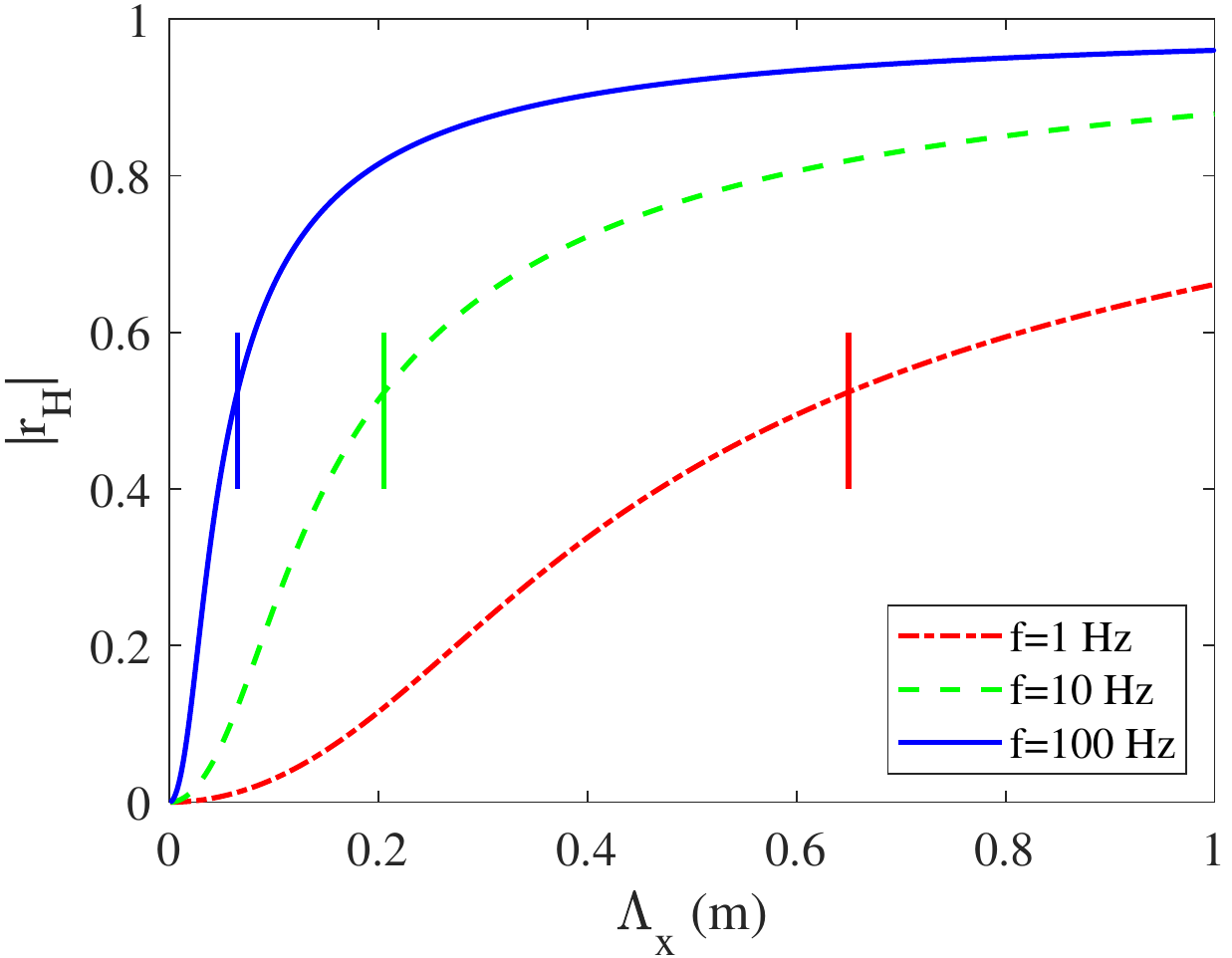} 
\caption{The reflection coefficient \eqref{rHqs} as a function of characteristic size $\Lambda_x=2\pi/k_x$ of the source for three different frequencies. The vertical lines indicate ten times the value of the skin depth, $10\delta$. The medium is Cu.}\label{fig:rkx} 
\end{center}
\end{figure}

By moving or changing the magnetic source, although slowly, there will be induced eddy currents in the medium. We now analyze this magnetic response with Fresnel's equations. Let medium 1 be vacuum ($\epsilon_1=\epsilon_0$), and medium 2 be a nonmagnetic medium with conductivity $\sigma$ described in the form of a complex, relative permittivity $\epsilon_2/\epsilon_0=1+i\sigma/\epsilon_0\omega\approx i\sigma/\epsilon_0\omega$. The skin depth is 
\be
\delta=\sqrt{2/\omega\mu_0\sigma}.
\ee
We consider a source which is uniform (and infinite) in the $y$-direction, with bound or free currents in the $\pm\vekh y$-direction. For example, the source may be a time-varying ``permanent'' magnet with magnetization in the $\vekh z$-direction (see Fig. \ref{fig:barmagnet}).
Then the magnetic field is independent of $y$, and has only $\vekh x$ and $\vekh z$ components. This leads to an electric field pointing in the $\vekh y$-direction (TE). The quantities we need in Fresnel's equations are
\begin{subequations}
\begin{align}
k_{1z} &= \sqrt{\omega^2/c^2-k_x^2}, \\
k_{2z} &= \sqrt{i\omega\mu_0\sigma-k_x^2}=\sqrt{i2/\delta^2-k_x^2}. \label{k2zqs}
\end{align}
\end{subequations}
Since the medium is passive, the sign of the square root is taken such that $\im k_{2z}> 0$. The magnetic field for $z<0$ becomes
\begin{align}
\vek H(x,z) &= \vekh x\int H_{\text{i}x}(k_x)\left[\e{ik_{1z}z}+r_H\e{-ik_{1z}z}\right]\e{ik_xx}\diff k_x \nonumber\\
&+ \vekh z\int H_{\text{i}z}(k_x)\left[\e{ik_{1z}z}-r_H\e{-ik_{1z}z}\right]\e{ik_xx}\diff k_x, \label{HTExqszn}
\end{align}
for some input spectra\footnote{Although not important here, these spectra are related by $k_xH_{\text{i}x}+k_{1z}H_{\text{i}z}=0$.} $H_{\text{i}x}(k_x)$ and $H_{\text{i}z}(k_x)$, and reflection coefficient
\be\label{rHqs}
r_H \equiv r^\TE_{H_x}=\frac{k_{2z}-k_{1z}}{k_{2z}+k_{1z}}.
\ee
In other words, we can still use an image source to create the field for $z<0$; however its field is filtered in the wavenumber domain using $r_H=r_H(k_x)$. This filter response is plotted in Fig. \ref{fig:rkx} for Cu ($\sigma\approx 6\cdot 10^7$\,S/m) for three different frequencies. We observe that the fine details (small $\Lambda_x$) are filtered away, such that the image gets smoother than the source. As seen from the figure, the cutoff size is around $10\delta$.

For $\Lambda_x\ll\delta\ll\lambda$ (where $\lambda$ is the free-space wavelength), we obtain
\be
r_H = -i\frac{1}{2k_x^2\delta^2}.
\ee
Using the permeability of Cu, we find that for $\Lambda_x=1$\,cm the magnetic response due to eddy currents is larger than the magnetic response due to the permeability, as long as $f>0.01$ Hz. Thus, the permeability of Cu ceases to describe the magnetic response in this setup already at the frequency 0.01~Hz.

Eq. \eqref{HTExqszn} expresses the magnetic field due to the source (first term in the brackets) and due to the resulting, induced currents in the medium (second term in the brackets). Thus \eqref{HTExqszn}-\eqref{rHqs} describe the interaction between the source and the medium. The reflection coefficient \eqref{rHqs} quantifies the strength of the interaction (Fig. \ref{fig:rkx}).

Similarly to \eqref{HTExqszn}, the magnetic field for $z>0$ can be expressed
\begin{align}
\vek H(x,z) &= \vekh x\int H_{\text{i}x}(k_x)t_H\e{ik_{2z}z+ik_xx}\diff k_x \nonumber\\
&+ \vekh z\int H_{\text{i}z}(k_x)t_H\frac{k_{1z}}{k_{2z}}\e{ik_{2z}z+ik_xx}\diff k_x,
\end{align}
for the transmission coefficient
\be\label{tHqs}
t_H \equiv t^\TE_{H_x}=\frac{2k_{2z}}{k_{2z}+k_{1z}}.
\ee

\section{Moving medium}\label{sec:moving}
We finally consider the case where medium 2 moves with velocity $v$ in the $x$-direction, as seen from the lab system. The coordinates of this lab system are shown in Fig. \ref{fig:setup}. We assume that medium 1 is vacuum ($\epsilon_1=\epsilon_0$ and $\mu_1=\mu_0$), and medium 2 is characterized by $\epsilon_2=\epsilon_\text{r}\epsilon_0$ and $\mu_2=\mu_\text{r}\mu_0$, where $\epsilon_\text{r}$ and $\mu_\text{r}$ are the relative permittivity and permeability, respectively. The source may be electrostatic (fixed charges) or magnetostatic (time-independent currents or permanent magnet), and is fixed in the lab system. We are interested in the reflection coefficients and coupling between electric and magnetic fields due to the moving medium 2. For the case with a magnetic source and a moving conducting medium, this response occurs as a result of eddy currents being induced in the conducting medium. We then obtain a connection to the analysis in Sec. \ref{sec:quasistatics}. In certain cases we will find that the response can be significant even for low (nonrelativistic) velocities.

We consider each spatial Fourier component of the source separately (with a certain $k_x$ and $k_y$), and  write the static electric and magnetic fields for $-d<z<0$ (see \eqref{EABz}):
\begin{subequations}\label{EHS}
	\begin{align}
	\vek E &= (k_x,k_y,k_z)U\e{ik_xx+ik_yy+ik_zz} \nonumber\\
	&+ (k_x,k_y,-k_z)V\e{ik_xx+ik_yy-ik_zz}, \\
	\vek H &= (k_x,k_y,k_z)S\e{ik_xx+ik_yy+ik_zz} \nonumber\\
	&+ (k_x,k_y,-k_z)T\e{ik_xx+ik_yy-ik_zz},
	\end{align}
\end{subequations}
where $k_z=i\sqrt{k_x^2+k_y^2}$. For an electrostatic source we are given $U$ and $S=0$, and want to find $r_E = V/U$ (reflection coefficient) and $c_{\text{\it EH}} = T/U$ (coupling). For a magnetostatic source we are given $S$ and $U=0$, and will find $r_H = T/S$ (reflection coefficient) and $c_{\text{\it HE}} = V/S$ (coupling). The strategy will be to transform \eqref{EHS} to the rest system of medium 2 using the usual relativistic transformation formulas \cite{landau_lifshitz_ctf}
\begin{subequations}\label{EHrel}
	\begin{align}
	E_x' &= E_x, \\
	E_y' &= \gamma E_y - \gamma v B_z, \\
	E_z' &= \gamma E_z + \gamma v B_y, \\
	B_x' &= B_x, \\
	B_y' &= \gamma B_y + \frac{\gamma v}{c^2} E_z, \\
	B_z' &= \gamma B_z - \frac{\gamma v}{c^2} E_y,
	\end{align}
\end{subequations}
where $\gamma = 1/\sqrt{1-v^2/c^2}$. We then use the electromagnetic boundary conditions to match the fields \eqref{EHrel} with those in medium 2.

In general we have that $(\omega/c,\vek k)$ is a four-vector, such that \cite{landau_lifshitz_ctf}
\begin{subequations}
	\begin{align}
	\frac{\omega'}{c} &= \gamma\left(\frac{\omega}{c} - \frac{v}{c} k_x\right), \\
	k_x'    &= \gamma\left(k_x-\frac{v}{c}\frac{\omega}{c}\right).
	\end{align}
\end{subequations}
This follows from the fact that phase is scalar: 
\be
\e{i\vek k\cdot\vek r-i\omega t}=\e{i\vek k'\cdot\vek r'-i\omega' t'}.
\ee
Here $t'$ and $\vek r'=(x',y,z)$ are the time and space coordinates in the rest system of medium 2, and $\omega'$ is the frequency in this system. Furthermore, $\vek k=(k_x,k_y,k_z)$ and $\vek k'=(k_x',k_y,k_z)$ are the wavevectors in the lab system and the rest system of medium 2, respectively. In our case we have static conditions ($\omega=0$) in the lab system, which means that 
\begin{subequations}\label{kkpwwp}
	\begin{align}
	\omega' &= -\gamma v k_x, \\
	k_x' &= \gamma k_x, \\
	\e{ik_xx} &=\e{ik_x'x'-i\omega' t'}.
	\end{align}
\end{subequations}

For simplicity, we will now consider two special cases with a two-dimensional source (which is homogeneous along the third direction).
\subsection{Case $k_x=0$ and $k_y\neq 0$}
Consider first the case where the source is homogeneous in the $x$-direction. In this case the source produces only components with  $k_x=0$. Then $\omega'=\omega=0$ and $\vek k'=\vek k=(0,k_y,k_z)$. The fields are therefore static and longitudinal, even in the rest system of medium 2. For $z>0$ we can write
\begin{subequations}\label{EHSzp2}
	\begin{align}
	\vek E' &= \left(0,k_y,k_z\right)F \e{ik_yy+ik_zz}, \\
	\vek H' &= \left(0,k_y,k_z\right)G \e{ik_yy+ik_zz}, 
	\end{align}
\end{subequations}
where $k_z = i|k_y|$, and $F$ and $G$ are arbitrary functions of $k_y$. Matching fields \eqref{EHrel} and \eqref{EHSzp2} according to the Maxwell boundary conditions, we obtain the reflection and coupling coefficient
\begin{subequations}\label{rckyEg}
	\begin{align}
	r_E \equiv \frac{V}{U} &= \gamma^2\left(\frac{1-\epsilon_\text{r}}{1+\epsilon_\text{r}}+\frac{v^2}{c^2}\frac{1-\mu_\text{r}}{1+\mu_\text{r}}\right), \label{rckyEgr}\\
	c_{\text{\it EH}} \equiv \frac{T}{U} &= i\,\text{sgn}(k_y)\gamma^2\frac{v}{c}\frac{1}{\eta}\left(\frac{1-\epsilon_\text{r}}{1+\epsilon_\text{r}}+\frac{1-\mu_\text{r}}{1+\mu_\text{r}}\right),	\label{rckyEgc} \end{align}
\end{subequations}
for the electrostatic source ($U\neq 0$, $S=0$). For the magnetostatic source ($U=0$, $S\neq 0$),
\begin{subequations}\label{rckyHg}
	\begin{align}
	r_H \equiv \frac{T}{S} &= \gamma^2\left(\frac{v^2}{c^2}\frac{1-\epsilon_\text{r}}{1+\epsilon_\text{r}}+\frac{1-\mu_\text{r}}{1+\mu_\text{r}}\right), \label{rckyHgr} \\
	c_{\text{\it HE}} \equiv \frac{V}{S} &= -i\,\text{sgn}(k_y)\gamma^2\frac{v}{c}\eta\left(\frac{1-\epsilon_\text{r}}{1+\epsilon_\text{r}}+\frac{1-\mu_\text{r}}{1+\mu_\text{r}}\right).	\label{rckyHgc} \end{align}
\end{subequations}
We have defined $\eta=\sqrt{\mu_0/\epsilon_0}$ as the vacuum wave impedance. In \eqref{rckyEg}-\eqref{rckyHg} the relative permittivity and permeability must be evaluated at zero frequency.

The reflections \eqref{rckyEgr} and \eqref{rckyHgr} can be interpreted as being produced by electrostatic and magnetic image sources, respectively. This interpretation corresponds exactly to that in Sec. \ref{sec:statics} except that the reflection coefficients $r_E$ and $r_H$ now are given by \eqref{rckyEgr} and \eqref{rckyHgr}. The differences compared to the situation with $v=0$ are of order $v^2/c^2$ and therefore negligible unless $v$ is a relativistic velocity.

However, for an electrostatic source, we also get a ``reflected'' magnetic field, and for a magnetostatic source, we also get a ``reflected'' electric field. For the case with a magnetostatic source, the resulting electric field needs not be negligible in the non-relativistic case. Assuming the parenthesis in \eqref{rckyHgc} is $\sim 1$, the electric field becomes $v\eta/c$ times the source magnetic field. This corresponds to an electric field strength $E\sim vB$, where $B$ is the flux density from the source. For a neodymium magnet with $B\sim 1\,$T and a velocity $v=1$~m/s, we get $E\sim 1\,$V/m.

The coupled fields \eqref{rckyEgc} and \eqref{rckyHgc} can also be described by image sources. Obviously, the needed image of an electric source will be magnetic, and vice versa. In addition the image will be distorted due to the $-i\,\text{sgn}(k_y)$ function. This filtering operation amounts to convolving by $1/\pi y$, which results in a Hilbert transform of the image with respect to $y$.

\subsection{Case $k_x\neq 0$ and $k_y=0$}
Consider next the case where the source is homogeneous in the $y$-direction, i.e., $k_y=0$. For a nonzero $k_x$, we observe from \eqref{kkpwwp} that $\omega'\neq 0$, despite the source being static in the lab system. Thus the electromagnetic field for $z>0$ in the rest system of medium 2 will be transverse, and can be expressed conventionally as a superposition of TE and TM modes: 
\begin{subequations}\label{EHSzp}
	\begin{align}
	\vek E' &= \left[ \left(0,1,0\right)F + \left(k_z',0,-k_x'\right)\frac{G}{\epsilon_\text{r}\epsilon_0\omega'} \right]\e{i\vek k'\cdot\vek r'-i\omega't'}, \\
	\vek H' &= \left[ \left(0,1,0\right)G + \left(-k_z',0,k_x'\right)\frac{F }{\mu_\text{r}\mu_0\omega'} \right]\e{i\vek k'\cdot\vek r'-i\omega't'}. 
	\end{align}
\end{subequations}
Here $F$ and $G$ are arbitrary functions of $k_x$. From the dispersion relation we have
\be
k_z' = \sqrt{\epsilon_\text{r}\mu_\text{r}\omega'^2/c^2 - k_x'^2}
	 = \gamma i|k_x|\sqrt{1-\epsilon_\text{r}\mu_\text{r} v^2/c^2},
\ee
where we have used \eqref{kkpwwp}. The sign of the square root is taken such that $\im k_z'>0$. Note that for dispersive media, we must evaluate the relative permittivity and permeability at the frequency $\omega'=-\gamma v k_x$. When this frequency is negative, the symmetry relations $\epsilon_\text{r}(-\omega')=\epsilon_\text{r}^*(\omega')$ and $\mu_\text{r}(-\omega')=\mu_\text{r}^*(\omega')$ are used \cite{landau_lifshitz_edcm}.

By matching the tangential electric and magnetic fields of \eqref{EHrel} and \eqref{EHSzp} at the boundary $z=0$, and using \eqref{kkpwwp}, we obtain the reflection coefficients
\begin{subequations}\label{rckx}
	\begin{align}
	r_E \equiv \frac{V}{U} &= \frac{\gamma\sqrt{1-\epsilon_\text{r}\mu_\text{r}v^2/c^2}-\epsilon_\text{r}}{\gamma\sqrt{1-\epsilon_\text{r}\mu_\text{r}v^2/c^2}+\epsilon_\text{r}}, \\
	r_H \equiv \frac{T}{S} &= \frac{\gamma\sqrt{1-\epsilon_\text{r}\mu_\text{r}v^2/c^2}-\mu_\text{r}}{\gamma\sqrt{1-\epsilon_\text{r}\mu_\text{r}v^2/c^2}+\mu_\text{r}}, \label{rckxm}
	\end{align}
\end{subequations}
where the relative permittivity $\epsilon_\text{r}$ and permeability $\mu_\text{r}$ must be evaluated at the frequency $\omega'=-\gamma v k_x$. The coupling coefficients between electric and magnetic fields in the lab system vanish in this case.

For a nondispersive medium 2, the reflection coefficients \eqref{rckx} are independent of $k_x$. Thus the field for $z<0$ can be seen as a superposition of that from the source and from an image multiplied by the reflection coefficient, exactly as in Sec. \ref{sec:statics}.

Of course, the most interesting regime is the nonrelativistic limit $v/c\ll 1$. Then the reflection coefficients reduce to the $v=0$ results in Sec. \ref{sec:statics}, unless the permittivity and/or permeability are extremely large for $\omega'=-\gamma v k_x$. However, this is exactly what happens for a conductor, for which $\epsilon_\text{r}(\omega') \approx i\sigma/\epsilon_0\omega'$ and $\mu_\text{r}=1$. In this case, \eqref{rckxm} can be written
\be\label{rmmet}
r_H = \frac{\sqrt{1+i\mu_0\sigma v/k_x}-1}{\sqrt{1+i\mu_0\sigma v/k_x}+1}.
\ee
Thus we obtain a considerable reflection when the dimensionless, magnetic Reynold's number $\mu_0\sigma v/k_x\gtrsim 1$. For Cu and  $v=1\,\text{m}/\text{s}$, this corresponds to spatial variations of the source which is $2\pi/k_x \gtrsim 0.1\,\text{m}$.

Intuitively, the reflection coefficient \eqref{rmmet} arises because there is a time-dependent magnetic field in the conducting medium, as seen from the rest frame of the medium. Through Faraday's law, eddy currents are induced, producing a ``reflected field''. Along these lines, \eqref{rmmet} can be found by considering the source as time-dependent (or quasistatic), as a result of a velocity $-v$ in the $\vekh x$-direction. Then we can use the result in Sec. \ref{sec:quasistatics} together with $\omega=-vk_x$ to deduce \eqref{rmmet} in the nonrelativistic limit. Despite giving the correct answer, such a method is somewhat incomplete unless the relativistic transformation formulas \eqref{EHrel} are taken properly into account.

\begin{figure}[]
	\begin{center}
		\includegraphics[width=8.3cm]{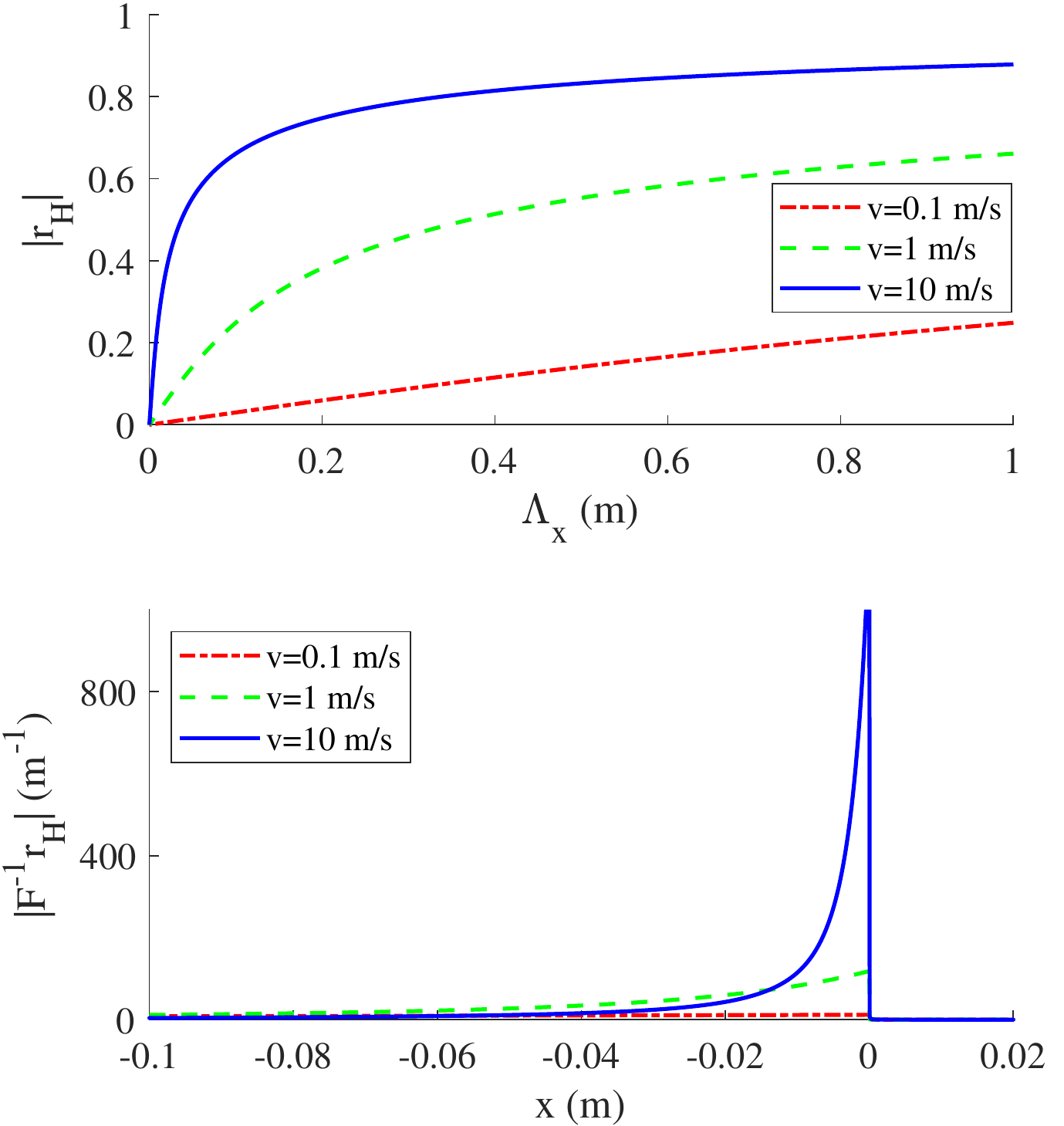} 
		\caption{The reflection coefficient $r_H=r_H(k_x)$ \eqref{rmmet} as a function of characteristic size of the source $\Lambda_x=2\pi/k_x$ (upper plot), and the inverse Fourier transform of $r_H(k_x)$ (lower plot).} \label{fig:filtermov} 
	\end{center}
\end{figure}

In the special case when the source is large and varies slowly with $x$, such that $\mu_0\sigma v/k_x \gg 1$ for the involved spatial frequencies, we obtain $r_H=1$. Then the image method from Sec. \ref{sec:statics} applies straightforwardly. In general, however, we must filter the image of the source with $r_H=r_H(k_x)$. This filter function is plotted with respect to characteristic size $\Lambda_x=2\pi/k_x$ in Fig. \ref{fig:filtermov}. Also the inverse Fourier transformed filter function (point source image) is plotted with respect to $x$. We observe the interesting result that the point source image is one-sided (i.e., vanishes for $x>0$). This results from the fact that \eqref{rmmet} can be analytically continued to the entire upper half-plane of complex $k_x$, since the branch point is in the lower half-plane. By closing the integration path with a large semi-circle in the upper half-plane, the inverse Fourier transform can therefore be proved to vanish for positive $x$. We finally note that since the image is not an even function, the system does not obey Lorentz reciprocity \cite{prat-camps2018}.

\section{Conclusion}
By formulating the proof of Fresnel's equations with only electric fields, or with only magnetic fields, it turns out that the relations are valid in statics. Since static fields are longitudinal, ``TM'' applies to the case with an electrostatic source, while ``TE'' applies to the case with a magnetostatic source. The results are independent of the spatial frequency of the source, which gives a connection to the classical image charge or image current methods. 

In quasistatics the Fresnel equations may give results dependent on spatial frequency; in this case the image method must be combined with a filter. For the case where a static source is located in the vicinity of a moving medium, we also obtain image methods to describe the reflection and coupling between electric and magnetic fields. Dependent on the orientation of the source relative to the velocity of the medium, certain interesting effects arise, including Hilbert transform of image, and Lorentz nonreciprocity.

\appendix
\section{Plane wave expansion of static fields}
We consider a general 3D charge distribution as the source, located at least a distance $d$ from the interface. In electrostatics the magnetic field is zero or constant, and we express $\vek E=-\grad\phi$, where $\nabla^2\phi=0$ away from the source and the interface. For each $z$, $\phi$ viewed as a function of $x$ and $y$ can be expressed as a 2D Fourier integral:
\be\label{phigeneral}
\phi(x,y,z) = \iint \Phi(k_x,k_y,z)\e{ik_xx+ik_yy}\diff k_x\diff k_y,
\ee
where $\Phi(k_x,k_y,z)$ is some function. Using $\nabla^2\phi=0$ we obtain
\be
\left(-k_x^2-k_y^2+\frac{\diff^2}{\diff z^2}\right)\Phi(k_x,k_y,z)=0,
\ee
with general solution
\be
\label{gensolAB}
\Phi(k_x,k_y,z)=iU(k_x,k_y)\e{ik_z z}+iV(k_x,k_y)\e{-ik_zz}.
\ee
Here $U$ and $V$ are arbitrary functions of $k_x$ and $k_y$. We have factored out an $i$ for later convenience, and defined
\be
k_z = i\sqrt{k_x^2+k_y^2}.
\ee
Backsubstitution into \eqref{phigeneral} we have
\begin{align}
\phi = i\iint &\left[U(k_x,k_y)\e{ik_zz}+V(k_x,k_y)\e{-ik_zz}\right] \nonumber\\
 & \cdot\e{ik_xx+ik_yy}\diff k_x\diff k_y.
\label{phiABz}
\end{align}
The associated electric field $\vek E=-\grad\phi$ becomes
\begin{align}
\vek E &= \iint (k_x,k_y,k_z)U(k_x,k_y)\e{ik_xx+ik_yy+ik_zz}\diff k_x\diff k_y \nonumber\\
 &+\iint(k_x,k_y,-k_z)V(k_x,k_y)\e{ik_xx+ik_yy-ik_zz}\diff k_x\diff k_y.
\label{EABz}
\end{align}
The result \eqref{EABz} is in particular valid in the region $-d<z<0$. For the unbounded region $z>0$ we can repeat the argument, introducing new integration constants, but the second term in \eqref{gensolAB} must be put to zero as it diverges with $z$. The result is
\be
\phi = i\iint W(k_x,k_y)\e{ik_xx+ik_yy+ik_zz} \diff k_x\diff k_y
\ee
and
\be
\vek E = \iint (k_x,k_y,k_z)W(k_x,k_y)\e{ik_xx+ik_yy+ik_zz} \diff k_x\diff k_y
\ee
for some function $W(k_x,k_y)$. Apparently there are no transverse fields in electrostatics.

If the source is a static current distribution or permanent magnet, we have $\curl H=0$ and $\div B=0$ away from the source. Thus we can express $\vek H=-\grad\psi$, where $\nabla^2\psi=0$, completely analogous to the electrostatic situation.

\def\cprime{$'$}
%


\end{document}